\documentclass[12pt,preprint]{aastex}

\newcommand{\srccshort}{SDSS J0836+0054} 
\newcommand{\srcbshort}{SDSS J1030+0524} 
\newcommand{\srcashort}{SDSS J1044-0125} 
\newcommand{\lrest}{\lambda_{\rm rest}} 
\newcommand{\Ol}{\Omega_\Lambda} 
\newcommand{\Om}{\Omega_m} 
\newcommand{\Hnot}{71\,km\,s$^{-1}$\,Mpc$^{-1}$}

\slugcomment{To appear in ApJ Letters }

\shorttitle{Fe in $z\approx6$ QSOs}
\shortauthors{Freudling et al.}

\begin{document}

\title{ Iron Emission in $z\approx6$ QSOs\footnote{Based on 
observations made with the NASA/ESA Hubble Space Telescope, obtained
at the Space Telescope Science Institute, which is operated by the
Association of Universities for Research in Astronomy, Inc., under
NASA contract NAS 5-26555. These observations are associated with
proposal \#9413.  }}

\author{Wolfram Freudling}
\affil{Space Telescope -- European Coordinating Facility \\
       European Southern Observatory \\
       Karl-Schwarzschild-Str. 2 \\
       85748 Garching \\
       Germany }
\email{wfreudli@eso.org}

\author{Michael R. Corbin}
\affil{Science Programs, Computer Sciences Corporation\\
       Space Telescope Science Institute\\
       3700 San Martin Dr.\\
       Baltimore, MD 21218}
\email{corbin@stsci.edu}

\and

\author{Kirk T. Korista}
\affil{Western Michigan University\\
      Physics Department\\
      1120 Everett Tower \\
      Kalamazoo, MI 49008-5252}
\email{kirk.korista@wmich.edu}

\begin{abstract}   

We have obtained low-resolution near infrared spectra of three QSOs at
$5.7 < z < 6.3$ using the NICMOS instrument of the Hubble Space
Telescope.  The spectra cover the rest-frame ultraviolet emission of
the objects between $\lrest\approx1600$\,\AA\, - $2800$\,\AA.  The Fe~II
emission-line complex at 2500\,\AA\ is clearly detected in two of the
objects, and possibly detected in the third.  The strength of this
complex and the ratio of its integrated flux to that of
Mg~II~$\lambda2800$ are comparable to values measured for QSOs at
lower redshifts, and are consistent with Fe/Mg abundance ratios near
or above the solar value. There thus appears to be no evolution of QSO
metallicity to $z\approx6$. Our results suggest that massive,
chemically enriched galaxies formed within 1~Gyr of the Big Bang. If
this chemical enrichment was produced by Type Ia supernovae, then the
progenitor stars formed at $z\approx20\pm10$, in agreement with recent
estimates based on the cosmic microwave background. These results also
support models of an evolutionary link between star formation, the
growth of supermassive black holes and nuclear activity.

\end{abstract}

\keywords{cosmology: observations --- 
          quasars:   emission lines --- 
          quasars:   individual (SDSS~J083643.85 + 005453.3, 
                                 SDSS~J103027.10 + 052455.0, 
                                 SDSS~J104433.04 - 012502.2) }

\section{Introduction}

The metallicities of high-redshift QSOs measured from their broad
emission lines are an important probe of the young universe.  These
metallicities can be used to infer the properties of the QSO host
galaxies, and to set limits on the formation of the first stars and
supernovae (see the review by Hamann \& Ferland 1999; Iwamuro et
al. 2002, Dietrich et~al.\ 2003).  The recent discovery of QSOs at
$z\approx6$ (Fan et~al.\ 2001; Fan et~al.\ 2003) is of particular
interest, as under currently favored cosmologies the age of the
universe at such redshifts is approximately 1\,Gyr, or only
$\approx$7\% of its current value. Such an age approaches the
formation timescale of the Type Ia supernovae that are the most likely
source of QSO iron enrichment (see Hamann \& Ferland 1999). Therefore,
QSOs at $z\approx6$ may begin to show significantly lower iron
abundances than those at lower redshifts.

With these motivations, we have begun to obtain near-infrared spectra
of QSOs at $z\approx6$ discovered from the Sloan Digital Sky Survey
(see Fan et~al.\ 2001, 2003 and references therein) using the Near
Infrared Camera and Multi-Object Spectrometer (NICMOS) of the Hubble
Space Telescope.  These NICMOS spectra cover a wider wavelength range
than is possible from the ground because of telluric absorption.  We
are particularly interested in the measurement of the strength of the
Fe~II emission complex centered at a laboratory wavelength of
2500\,\AA, and the ratio of this emission to that of Mg~II
$\lambda$2800, as this provides a diagnostic of the abundance ratio
of iron to $\alpha$-process elements, which in turn is an indicator of
the occurrence of SNe~Ia (see Hamann \& Ferland 1999; Thompson et~al.\
1999; Iwamuro et~al.\ 2002).

In this {\it Letter} we report the first results from this program.  A
detailed discussion of our data reduction procedures and a full
modeling of the data are deferred to a later paper. Throughout this
paper, we assume a cosmology of $H_o$=\Hnot, $\Ol$=0.73, and
$\Om=0.27$ (Bennett et~al.\ 2003).

\section{Observations and Data Reduction}

Spectra of the three QSOs SDSS J083643.85+005453.3 ($z$ = 5.82), SDSS
J103027.10+052455.0 ($z$ = 6.28), and SDSS J104433.04-012502.2 ($z$ =
5.78) were obtained between 2002 October~25-27 using the G141 grism of
the NICMOS instrument on the Hubble Space Telescope under GO program
9413. Hereafter, we refer to these sources as \srcashort, \srcbshort\
and \srccshort. This grism covers the wavelength range from 1.1$\mu$
to 1.9$\mu$ with a nominal resolution of 200 per pixel. The final
spectra represent the average of four individual spectra centered in each
of the four NICMOS Camera 3 detector quadrants. Their position on the
detector was changed in order to minimize the effect of large-scale
and intra-pixel sensitivity variations in the detector on the co-added
spectra. The total integration time for each spectrum was 2304
seconds. The spectra were extracted using the NICMOSlook software of
Freudling (1997). Intra-pixel variations show up as a wavy pattern in
individual spectra with an amplitude of approximately 8\% of the mean
flux level. The effect was corrected for during the extraction. The
wavelength calibration of the spectra was derived from the
observations of the planetary nebular HB12, and the flux calibration
is based on the solar analog star P330-E (Freudling 2003) using a
spectrum kindly provided by Rieke (2001). {Galactic extinction at the
position of the QSOs is negligible and was not corrected for.}

\section{Results}

\begin{figure}
\plotone{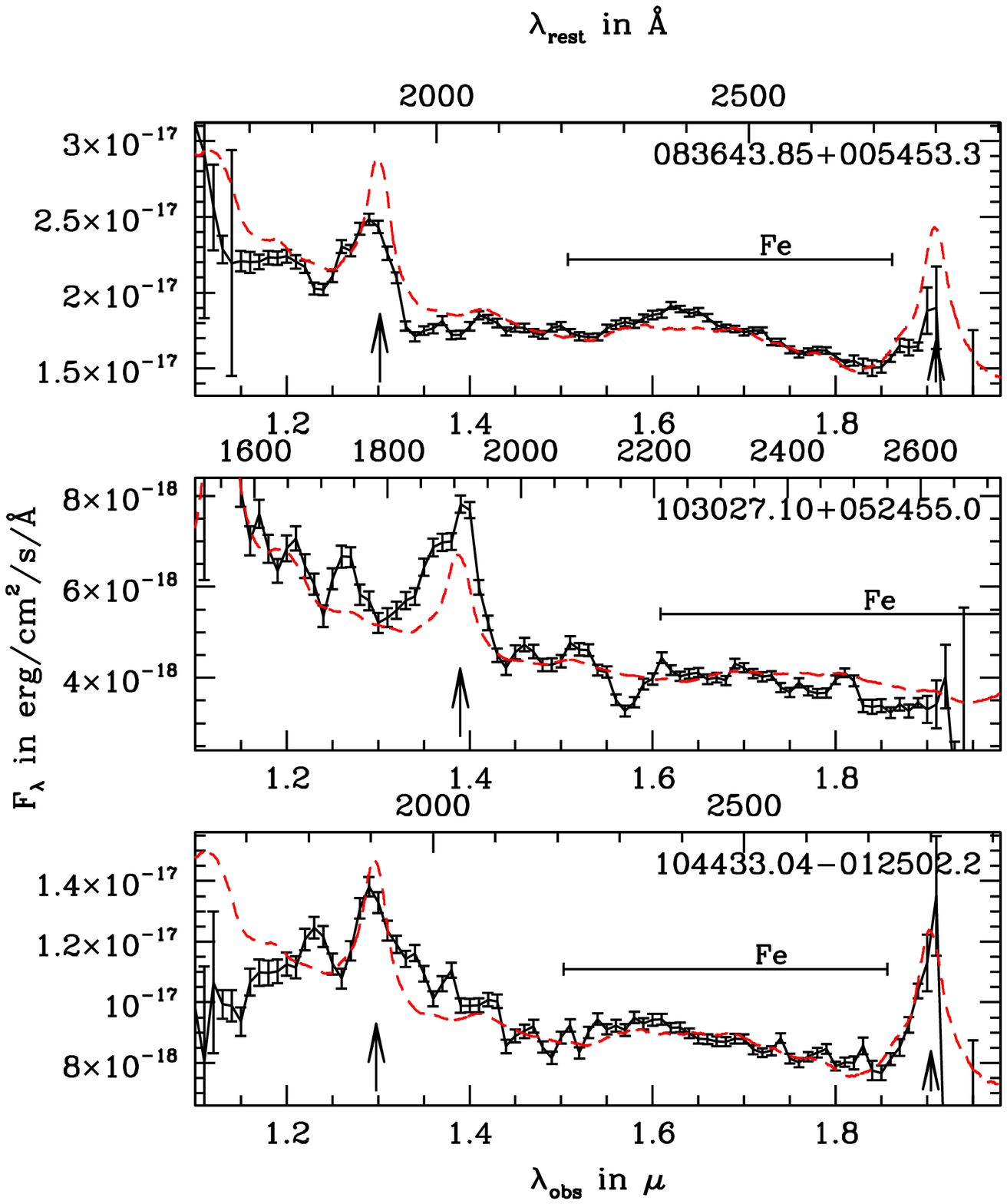}
\caption{NICMOS spectra of our sample QSOs. The connected points are the 
NICMOS data. The error bars are the instrumental random errors. The
wavelength range $2210 < \lambda_{\rm rest} < 2730$\,\AA\ is labeled
``Fe''. The expected positions of the C~III]~$\lambda1909$, and
Mg~II$\lambda2800$ lines are marked with arrows. The dashed lines show
the scaled and redshifted composite spectrum of Zheng et
al. (1997). }\label{spectra}
\end{figure}

Our final spectra are shown in Figure~\ref{spectra}, plotted as a
function of observed and rest-frame wavelengths.  The positions of the
C~III] $\lambda$1909 and Mg~II $\lambda$2800 lines are marked with
arrows, and the wavelength range of the main Fe~II UV emission bump is
marked with a broad bar. The noise in the spectra is slightly
correlated due to rebinning and flat-fielding uncertainties. The
signal-to-noise ratios of our spectra are about 60 for
SDSS~J08364+0054 and SDSS~J1044-0125, and about 30 for
SDSS~J1030+0524. 

Each spectrum in Figure~\ref{spectra} is overplotted with the QSO
composite spectrum of Zheng et~al.\ (1997) derived from HST Faint
Object Camera observations of objects in the approximate range $0.33 <
z < 1.5$, shifted to the appropriate redshift.  One can immediately
see a basic correspondence between the continuum shapes of all three
object spectra and that of the composite spectrum.  In SDSS J0836+0054
and SDSS J1044-0125 one can also see the presence of the main UV Fe~II
emission bump and its correspondence to this feature in the composite.
The detection of the Fe II emission bump in SDSS~J1044-0125 is
consistent with the detection of Fe II emission at $\lambda_{\rm
rest}\approx$ 3000\,\AA\ reported in this object by Aoki, Murayama \&
Denda (2002).

The spectrum of SDSS J1030+0524 is more ambiguous. The signal-to-noise
ratio of this spectrum is the lowest of the three objects, adding to this
uncertainty.  There is an apparent absorption feature at approximately
1.57$\mu$, which we also see in the four individual spectra, indicating
that it is not a detector artifact.  There are no objects nearby the
QSO whose spectra could have contaminated it.  We thus believe the feature
to be real.  It could represent a deep complex of Mg~II $\lambda$2800
narrow line absorbers at $z \approx$ 4.6, or else be intrinsic to the QSO.
The latter possibility might explain the apparent lack of strong emission
in the Fe~II complex, i.e., SDSS J1030+0524 might be related to the
rare sub-class of QSOs showing Fe~II absorption in their spectra (see
Gregg et al.\ 2002 and references therein). A spectrum with a higher
signal-to-noise ratio is required to resolve this uncertainty.

\section{ Measurement of the Fe~II Emission Complex and Comparison with
lower redshift QSOs}

\begin{figure}
\plotone{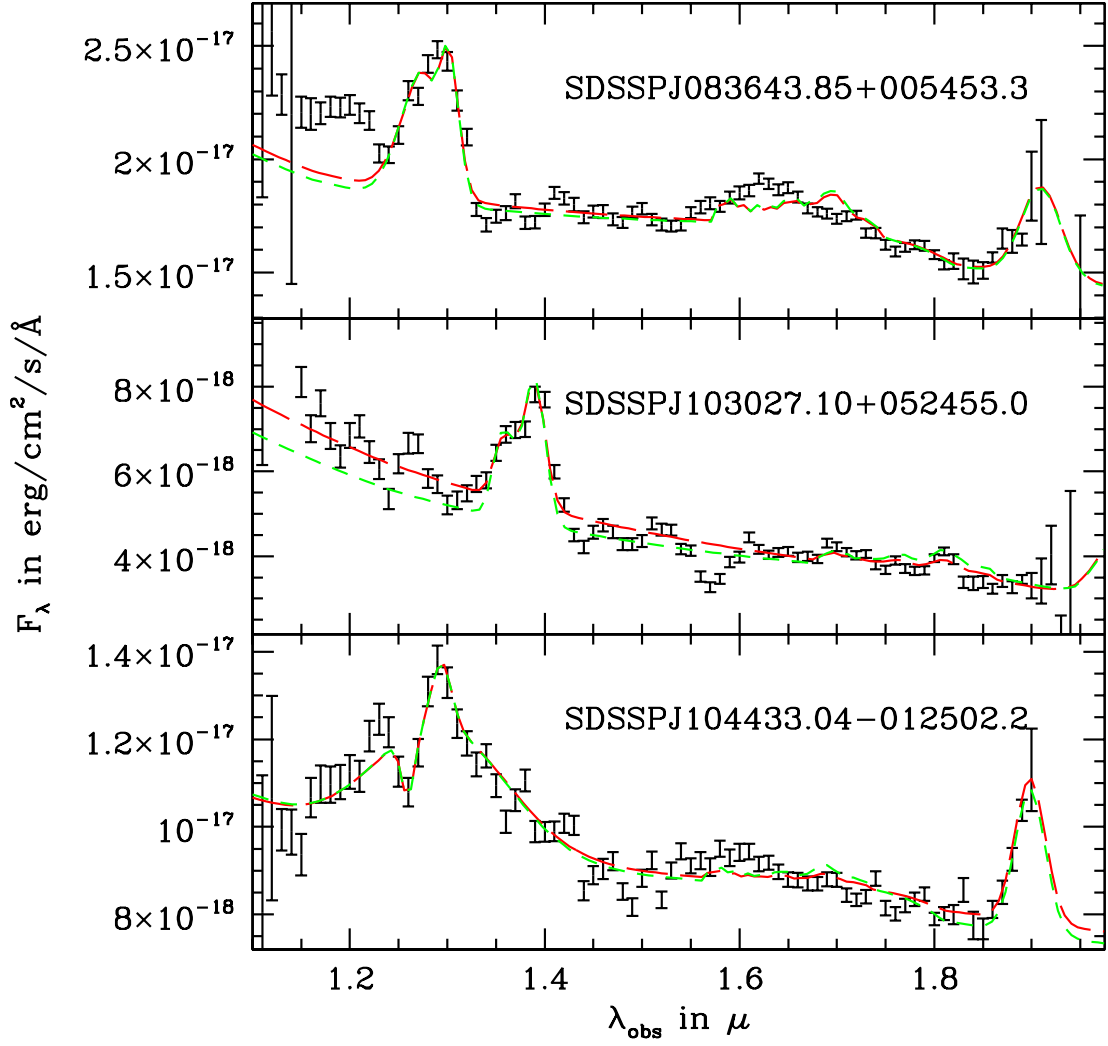}
\caption{
Fits of continuum, Fe~II emission, and Mg~II $\lambda2800$ and C~III]
$\lambda1909$ lines to our object spectra.  The long-dashed and
short-dashed lines show fits with different continuum slopes.
}
\label{fit}
\end{figure}

We measure the strength of the detected Fe II emission and its ratio
to that in Mg~II $\lambda$2800 as follows.  Following an approach
first used by Wills, Netzer \& Wills (1985; hereafter WNW85), we use a
smoothed version of the composite spectrum of Zheng et al. (1997) as a
template of the Fe~II emission complex in the region 2250\,\AA\ --
2700\,\AA\, and extend it to both longer and shorter wavelengths using
the theoretical Fe~II template from WNW85 as a guide.  We then model
our NICMOS spectra as the combination of a continuum, our UV Fe~II
template, and Gaussian profiles. The continuum in our model
consists of a power law with slope $\alpha = -0.33$ and a model Balmer
continuum. All components are fitted simultaneously. The Mg~II
$\lambda$2800 line in the QSO spectra is fitted by a single Gaussian
profile, and the $\lambda1900$ line blend that includes C~III]
$\lambda$1909, Si III] $\lambda$1892, and Al~III $\lambda$1860 is
fitted as two Gaussian profiles. In the case of SDSS J1030+0524, a
third broader Gaussian profile is added to this blend to accommodate
its shape. The redshift of the QSO is a free parameter in the fits,
the resulting values are given in table~1. Fits to each spectrum are
shown in Figure~\ref{fit}. While it can be seen from the plots that
Mg~II $\lambda$2800 is incompletely covered in the spectra of
SDSS~J08364+0054 and SDSS~J1044-0125, the shape of the line profiles
suggest that the peak of the lines are within or very close to the
wavelength range covered by our spectra.  Fits to the blue sides of
the profiles therefore allow a reasonable estimate of their total
flux. For SDSS J1030+0524 where Mg~II $\lambda$2800 is not covered, we
use the CIII] line blend instead and assume that F(Mg~II)$ =
1.37\cdot$F(CIII]). The factor 1.37 is the mean flux ratio in
radio-quiet QSOs (Brotherton et~al.\ 1994).

The main source of uncertainty in the measured F(Fe~II) is the value
of the continuum slope. We therefore repeat our fitting procedure with
different slopes between $\alpha=$-0.30 and -0.36. For each slope, we
judge by eye whether the fitted continuum corresponds to the measured
flux values within their error bars and discard those which do
not. Two of the remaining fits are shown for each object in
Figure~\ref{fit}.  We compute integrated fluxes F(Fe~II) and F(Mg~II)
for each of the accepted continuum fits. We then take half the value
of the difference between the maximum and minimum integrated flux as
our error estimate. We note that for all continuum fits accepted 
during the inspection described above,
we obtain a positive value for the integrated flux of the Fe~II
UV emission complex in SDSS J1030+0524. Nevertheless, because of the
aforementioned ambiguity in our spectrum of this object, this value
should only be considered an upper limit.

A more accurate measurement of the Fe~II / Mg~II ratio is possible in
principle through the use of a Fe~II template derived from the
narrow-line QSO I~Zw~1 (see Boroson \& Green 1992; Corbin \& Boroson
1996; Vestergaard \& Wilkes 2001), but given our low spectral
resolution, wavelength coverage and relatively low signal-to-noise
ratios it is not clear that our data warrant this approach. However,
we will test that method on our complete sample of spectra.

The integrated flux of the Fe~II $\lambda$2500 complex is generally
measured over the wavelength range from 2150 to 3300\,\AA\
(e.g. WNW85), and the continuum fit is performed over the wavelength
range from 1500 to 5000\,\AA. Our spectra however do not extend out
this far in the object rest-frame, and therefore we can only measure
the Fe~II emission at wavelengths shorter than the Mg~II line. In
order to compare our measurements of the F(Fe~II)/F(Mg~II) ratios to
measurements of the same quantity in QSOs at lower redshifts by
Iwamuro et~al.\ (2002), we computed a scaled version of our
Fe~II/Mg~II ratios,

\begin{equation}
[{\rm F(Fe~II)/F(Mg~II)}]_{\rm scaled}= s \times 
{ [{\rm F(Fe~II)/F(Mg~II)}]_{\rm measured} .  }
\end{equation}

The scaling factor $s$ was derived using our procedure to measure line
fluxes on the template spectrum by Francis et~al.\ (1991), for which
Iwamuro et~al.\ (2002) have reported a value for Fe~II/Mg~II of 4.26.
We measure in this spectrum the Fe~II/Mg~II flux ratio using the same
wavelength range, continuum regions, line and template fitting as for
our NICMOS spectra. The final factor scaling $s$ is chosen so that our
scaled Fe~II/Mg~II flux ratio for the Francis spectrum is identical to
the value quoted by Iwamuro et~al.  The measured and scaled values of
this ratio are given in Table~1. While our measurement of the scaled
Fe/Mg ratio in SDSS J1030+0524 is highly uncertain, a value near or
above where we measure it would be consistent with the results of
Pentericci et~al.\ (2002), who from an analysis of the lines present
in the optical spectrum of this object conclude that that its
metallicity is likely to be above the solar value.

\begin{figure}
\plotone{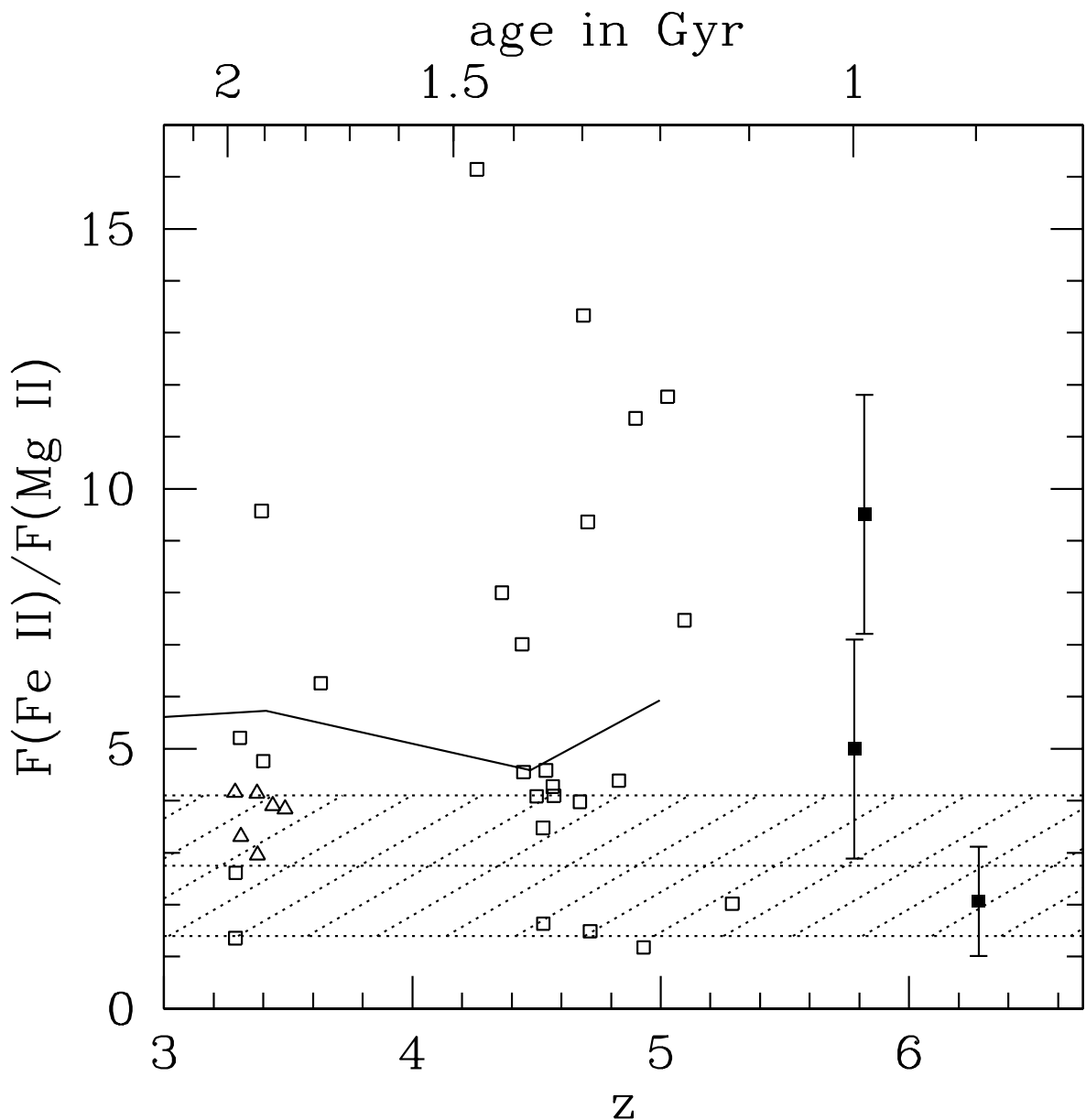}
\caption{Integrated Fe~II/Mg~II flux ratios  plotted as a function of 
redshift (lower axis) and age of the universe (upper axis). Open
squares are data from Iwamuro et~al.\ (2002), open triangles from
Dietrich et al (2002), and solid squares from this work.  The solid
line shows the averages in redshift bins given by Iwamuro et
al. (2002). The shaded area is the range of this ratio found by WNW85
to correspond to solar metallicity. }\label{fez}
\end{figure}

Our measurement procedure is similar to those used by WNW85, Iwamuro
et al. (2002) and Dietrich et al. (2002). Our scaling should therefore
allow our measurements to be compared with theirs.  In
Figure~\ref{fez} we plot our values in combination with the values
listed by Iwamuro et~al.\ (2002, their tables~2 and~3) and Dietrich
et~al. (2002, column ``Fe~II~UV'' of their table~4) as a function of
redshift. The shaded region in this plot indicates the flux ratio of
$2.75\pm1.35$ expected for solar abundances (see WNW85). The figure
shows that the measured F(Fe~II)/F(Mg~II) values are generally at or
above this range of values, and more importantly show no trend with
redshift over the range $z\approx 3-6.3$. All the QSOs in
this combined sample have comparable luminosities. Thus while Iwamuro
et~al.\ (2002) find evidence for some redshift dependence of the Fe/Mg
abundance ratio between high and low-redshift ( $z \approx$ 0.1) QSOs,
the interpretation of this trend is complicated by the weak dependence of
the Mg~II $\lambda$2800 equivalent width on continuum luminosity, and
the wide luminosity range of their sample. There is thus no compelling
evidence of evolution in QSO Fe/Mg abundance ratios from $z \approx 6$ to
$z \approx 0.1$.

\section{Discussion}

The evidence that the objects in our sample have Fe/Mg abundance ratios
at or above the solar value has several important implications.  First,
it supports models which predict that galaxy spheroids can reach solar
or super-solar metallicities within $\approx$ 1\,Gyr of their formation
(e.g.  Matteuchi 1994; Gnedin \& Ostriker 1997; Romano et~al.\ 2002).
Second, the relationship between mass and metallicity found among
local galaxies suggests that the hosts of these objects are very massive
($\sim10^{11}-10^{12} M_\sun$), possibly large elliptical galaxies (see
Hamann \& Ferland 1999).  The presence of such galaxies at $z\approx6$
would then support scenarios of an early formation epoch of massive
galaxies (e.g. Cowie et~al.\ 1996; Daddi, Cimatti \& Renzini 2000,
Romano et~al.\ 2002), as opposed to those in which they form via mergers at
lower redshifts (e.g. Zepf 1997; Barger et al. 1999).

If the high metallicities suggested in our sample objects were
produced by SNe Ia, then our assumed cosmology implies that these
supernovae occurred less than 1 Gyr after the Big Bang.  Their stellar
progenitors  must have evolved and gas must have sufficiently
mixed before high Fe~II/Mg~II flux ratios are detectable. Matteucci \&
Recchi (2001) have shown that the formation timescale of SNe~Ia is
strongly dependent on environment, initial mass function, and star
formation rate. Only several 100\,Myr after the maximum SNe~Ia rate
does the gas Fe/Mg abundance ratio reach its solar values (Matteucci
1994). Recent models indicate that the time scale from the formation
of SN~Ia progenitor stars until the gas has been enriched ranges from
0.5 to 0.8\,Gyr (Friaca \& Terlevich 1998).  The evidence of
enrichment from SNe Ia at $z\approx6$ thus indicates that the
progenitor stars formed at $z\approx20\pm10$. It is interesting to
note that this range is similar to the range of re-ionization
redshifts derived from the polarization fluctuations in the cosmic
microwave background (Bennett et~al.\ 2003). This would support the
suggestion that stars rather than QSOs were mainly responsible for
reionization (Yan, Windhorst \& Cohen 2002).

It is likely that QSOs at $z\approx6$ have only recently turned on.
This is indicated mainly by their extremely low space density (see
Shaver et~al.\ 1996; Fan et~al.\ 2003).  Haiman \& Cen (2002) conclude
from the size of the ionized volume that SDSS~J1030+052 has created in
the surrounding neutral intergalactic medium, that it has been
emitting at its observed luminosity for only $\sim2\times10^7$ years.
The evidence that star formation and supernovae began much earlier
than the nuclear activity in these objects thus argues in favor of
models proposing an evolutionary link between nuclear starbursts and
the AGN phenomenon (Norman \& Scoville 1988; Kauffmann \& Haehnelt
2000; Kawakatu \& Umemura 2003), as opposed to models in which the
formation of supermassive black holes precedes that of galaxies. In
the latter scenario, one would expect a more constant space density of
QSOs at high redshifts along with an evolution in QSO metallicity (see
Loeb 1993a; 1993b).

\acknowledgments

{\it Acknowledgments}.  We dedicate this paper to the memory of the
final crew of the Space Shuttle {\it Columbia}. The achievements of
the Hubble Space Telescope would not be possible without the efforts
of the dedicated astronauts of NASA, and we thank them for their
service over the last thirteen years.  We acknowledge useful
discussion with Peter Shaver, and helpful comments by the referee
Keith Thompson.

Support for proposal 9413 was provided by NASA through a grant from the
Space Telescope Science Institute, which is operated by the Association
of Universities for Research in Astronomy, Inc., under NASA contract
NAS5-26555.

\clearpage

\begin{deluxetable}{cccccc} 
\tablecolumns{6} 
\tablewidth{0pc} 
\tablecaption{Flux Ratios (Rest-Frame)} 
\tablehead{ 
\colhead{Source} & z\tablenotemark{a}         &  \colhead{F(Fe~II)/F(Mg~II)} & \colhead{F(Fe~II)/F(Mg~II)}   \\
\colhead{}       &\colhead{} &  \colhead{measured} & \colhead{scaled}\\
}
\startdata 
\srccshort & 5.82 &  5.41 $\pm$ 0.5                  & 9.5 $\pm$ 2.3   \\
\srcbshort & 6.28 &  1.18 $\pm$ 0.5\tablenotemark{b} & 2.1 $\pm$ 1.1  \\
\srcashort & 5.78 &  2.84 $\pm$ 1.0                  & 5.0 $\pm$ 2.1  \\
\enddata
\tablenotetext{a}{Redshift as determined from fit of emission lines}
\tablenotetext{b}{F(Mg~II) estimated from CIII] line blend}
\end{deluxetable}

\end{document}